\documentclass[a4paper,prd,showpacs,noshowkeys,preprint]{revtex4}
\usepackage[latin1]{inputenc}
\usepackage{amsmath}
\usepackage{amssymb}
\usepackage{mathrsfs}

\providecommand{\bp}{{\bf p}}
\providecommand{\bk}{{\bf k}}
\providecommand{\bq}{{\bf q}}
\providecommand{\bn}{{\bf n}}
\providecommand{\hn}{\hat{\bf n}}
\providecommand{\ip}[1]{#1{\cdot}}
\providecommand{\tM}{\tilde{\mathcal{M}}}
\providecommand{\LL}{\overset{\,\leftarrow}{\leftarrow}}
\providecommand{\RR}{\overset{\rightarrow\,}{\rightarrow}}
\providecommand{\RL}{\overset{\,\rightarrow}{\leftarrow}}
\providecommand{\LR}{\overset{\leftarrow\,}{\rightarrow}}

\providecommand{\lag}{\mathcal{L}}
\providecommand{\sla}[1]{~\hs{-1.5ex}\not\hs{-.4ex}#1\hs{.1ex}}

\providecommand{\eq}[1]{\begin{equation} #1 \end{equation}}
\providecommand{\eqarr}[1]{\begin{eqnarray} #1 \end{eqnarray}}
\providecommand{\ket}[1]{\vert #1 \rangle}
\providecommand{\bra}[1]{\langle #1 \vert}
\providecommand{\hs}[1]{\hspace{#1}}
\providecommand{\mn}[1]{\mbox{\normalsize $#1$}}
\providecommand{\ms}[1]{\mbox{\small $#1$}}
\providecommand{\diag}{\mathrm{diag}}
\begin{document}
\title{Absolute neutrino mass from helicity measurements}
\author{C.~C.~Nishi
}
\email{ccnishi@ifi.unicamp.br}
\affiliation{
Institute of Physics ``Gleb Wataghin''\\
University of Campinas, UNICAMP\\
13083-970, Campinas, SP, Brazil
}
\affiliation{
Instituto de Física Teórica,
UNESP -- São Paulo State University\\
Rua Pamplona, 145,
01405-900 -- São Paulo, Brasil
}
\begin{abstract}
The possibility to access the absolute neutrino mass scale through the
measurement of the wrong helicity contribution of charged leptons is
investigated in pion decay. Through this method, one may have access to the
same effective mass $m^2_\beta$ extractable from the tritium beta decay
experiments for electron neutrinos as well as the analogous effective mass
$(m^2_{\nu_\mu})_{\rm eff}$ for muon neutrinos.
In the channel $\pi^-\rightarrow e^-\bar{\nu}$, the relative probability of
producing an antineutrino with left helicity is enhanced if compared with the
naive expectation $(m_\nu/2E_\nu)^2$. The possibility to constrain new
interactions in the context of Two-Higgs-Doublet models is also investigated.
\end{abstract}
\pacs{14.60.Pq, 13.15.+g, 13.20.Cz}
\maketitle
\section{Introduction}
\label{sec:intro}

After the confirmation that neutrinos have non-null masses and non trivial
mixing among the various types, the knowledge of the absolute scale of neutrino
masses is one of the most urgent questions in neutrino physics.
In recent times, the greatest advances in the understanding of neutrino
properties were boosted by neutrino oscillation experiments which are only
capable of accessing the two mass squared differences and, in principle,
three mixing angles and one Dirac CP violating phase of the
Maki-Nakagawa-Sakata (MNS) leptonic mixing matrix.

Too large masses for the light active neutrinos may alter significantly the
recent cosmological history of the Universe. Indeed, the most 
stringent bounds for the value of the sum of neutrino masses come from
Cosmology:
\eq{
\label{mnu<}
\sum_\nu m_\nu < 0.17 \mathrm{eV}
}
at 95\% confidence level\,\cite{seljak.06:nu}

Another issue of great interest refers to the existence of heavy right-handed
neutrinos which could explain at the same time the tiny active neutrinos masses
through the seesaw mechanism as well as the matter-antimatter asymmetry of
the Universe through the implementation of the mechanism of
bariogenesis through leptogenesis\,\cite{leptog}.

Despite of the stringent bound \eqref{mnu<} coming from cosmological analyses,
terrestrial direct search experiments establish much more looser
bounds\,\cite{pdg}:
\eqarr{
\label{mnue<}
m_{\nu_e}&\le& 2 \mathrm{eV}\,, \\
\label{mnumu<}
m_{\nu_\mu}&\le& 190 \mathrm{keV}\,, \\
\label{mnutau<}
m_{\nu_\tau}&\le& 18.2 \mathrm{MeV}\,.
}
These bounds are based on ingeniously planned experiments\,\cite{giunti}, but
their intrinsic difficulties rely on the fact that they should probe,
essentially, the kinematical effects of neutrino masses which are negligible
compared to other typical quantities involved in processes with neutrino
emission.

Although the bounds in Eqs.\,\eqref{mnue<}--\eqref{mnutau<} are not as
stringent as the cosmological bound in Eq.\,\eqref{mnu<}, it is always
desirable to have a direct measurement of neutrino masses. Two more reasons can
be listed in favor of direct terrestrial searches: (a) cosmological bounds may
be quite model dependent and (b) we may have access to mixing parameters through
the effective neutrino masses.
For the electron neutrino, there are ongoing experiments planning to reduce the
respective bound to 0.2 eV\,\cite{katrin}.

The main goal of this article is to investigate the possibility of accessing the
absolute neutrino mass scale through one of the most natural consequences of
massive fermions, i.e., the dissociation of \textit{chirality} and
\textit{helicity}. Consider the pion decay $\pi^-\rightarrow
\mu^-\bar{\nu}_\mu$. Since the pion is a spin zero particle, in its rest frame,
the decaying states should have the following form from angular momentum
conservation,
\eq{
\label{RR+LL}
\ket{\pi}\rightarrow\quad
\ket{\mu\!:\LL}\ket{\bar{\nu}\!:\RR}
+\delta \ket{\mu\!:\RL}\ket{\bar{\nu}\!:\LR}
\,,
}
where the arrows represent the momentum direction (longer arrow) and the spin
direction (shorter arrow); the combination $\RR$, for example, means a
rightgoing fermion with a right helicity. On later calculations 
the right (left) helicity will be denoted simply as h=+1 (h=-1).
The normalization of the state is arbitrary and the coefficient $\delta$ is of
the order of $m_\nu/E_\nu$, which will be calculated in Sec.\,\ref{sec:pidecay}.
For massless neutrinos there is no second term in Eq.\,\eqref{RR+LL} since the
antineutrino is strictly right-handed in helicity and chirality. Thus by
measuring the wrong helicity contribution of the charged lepton it is possible
to have access to the neutrino mass. 
Such possibility was already suggested in Refs.\,\cite{shrock,shrock2} but we intend here
a focused reanalysis of the possibility considering the present experimental
bounds.
The precision in the polarization measurement necessary to extract the wrong
helicity is also calculated while the possibility to constrain new
interactions in the context of the two-Higgs-doublet model (2HDM) is
investigated as well.

It is interesting to notice that there has been 50 years since the first
measurement of the helicity of the electron neutrino\,\cite{nuehelicity}. At
that time the primary concern was to confirm the V-A theory of weak
interactions. Nowadays, we can try to invert their roles to obtain new
information about the neutrinos from the well established weak interaction part
of the Standard Model (SM). Measurements of the muon neutrino and the
tau-neutrino helicities can be also found in the
literature\,\cite{roesch,fetscher,nutauhelicity}.

\section{Pion decay}
\label{sec:pidecay}

Pion decay $\pi^-\rightarrow l_i^-\bar{\nu}_j$ can be effectively described by
the four-point Fermi interaction Lagrangian
\eq{
\label{L:CC}
\lag_{\rm CC}=
2\sqrt{2}G_F \bar{l}_i\gamma^\mu LU_{ij}\nu_j J_\mu + h.c.,
}
where $L=\frac{1}{2}(1-\gamma_5)$, $\{U_{ij}\}$ denotes the MNS matrix while
$J_\mu$ is the hadronic current
that in the case of pion decay reads
\eq{
J_\mu=V_{ud} \bar{u}_L\gamma_\mu d_L
\,.
}

From Eq.\,\eqref{L:CC} it is straightforward to calculate the amplitude for
$\pi(\bp)\rightarrow l_i(\bq)\nu_j(\bk)$, $i,j=1,2$,
\eq{
\label{pi:amp:1}
-i\mathcal{M}(\pi\rightarrow l_i\bar{\nu}_j)=
2G_FF_\pi V_{ud}
\bar{u}_{l_i}(\bq)\sla{p}LU_{ij}v_{\nu_j}(\bk)
}
by using\,\cite{DGH}
\eq{
\bra{0}\bar{u}\gamma_5\gamma_\mu d\ket{\pi^-}=ip_\mu\sqrt{2} F_\pi
\,,
}
where $F_\pi\approx 92 \rm MeV$ is the pion decay constant.
Let us denote the spinor dependent amplitude as
\eq{
\tM_{ij}=
\bar{u}_{l_i}(\bq)\sla{p}LU_{ij}v_{\nu_j}(\bk)
\,.
}
\eqarr{
\label{M2:spins}
\sum_{\rm spins}|\tM_{ij}|^2
&=&
4(\ip{p}{q_i})(\ip{p}k_j)-2p^2(\ip{q_i}k_j)\\
&=&
\label{p=q+k}
M^2_i(M^2_\pi-M^2_i)
\cr&& +\ 
m^2_j(M^2_\pi+2M^2_i-m^2_j)
\,.
}
The last expression \eqref{p=q+k} is exact and follows when $p=q_i+k_j$
(four-vector), $p^2=M^2_\pi$, $q_i^2=M^2_i$ and $k_j^2=m^2_j$.
The first expression \eqref{M2:spins} does not assume energy-momentum
conservation.

We can calculate the amplitude squared, summed over the neutrino spin, but
dependent on the polarization $\hat{\bn}$ of the charged lepton in its rest
frame:
\eqarr{
\label{A^2:n}
P_{ij}(n_i)&\equiv&
\sum_{\nu_j \text{ spin}}|\tM_{ij}|^2
\\&=&
M^2_i[\ip{q_i}k_j+M_i(\ip{k_j}n_i)]
+ 2M^2_im^2_j
\cr&&
+\ m^2_j[\ip{q_i}k_j-M_i(\ip{k_j}n_i)]
\,,~~~
}
where
\eq{
n_i^\mu=\Big(\frac{\ip{\bq}\hn}{M_i},
\hn+(\frac{E_{l_i}}{M_i}-1)(\ip{\hn}\hat{\bq})\hat{\bq}\Big)
}
For the particular directions $\hn= h_i\hat{\bq}$ we single out the
positive ($h_i=1$) and negative ($h_i=-1$) helicity for the charged
lepton\,\cite{IZ}.

In the helicity basis
\eq{
\ip{q_i}k_j\pm M_i(\ip{k_j}n_i)=
[E_{l_i}(\bk)\pm h_i|\bq|][E_{\nu_j}\mp h_i\ip{\hat{\bq}}\bk]
}
Therefore, for the pion at rest we obtain
\eqarr{
P_{ij}(h_i=1)&=&
M^2_i(M^2_\pi-M^2_i) + O(m^2_j)
\\
P_{ij}(h_i=-1)&=&
m^2_j\frac{M^4_\pi}{M^2_\pi-M^2_i}+ O(m^4_j)
}
Without approximations we obtain
\eq{
P_{ij}(h_i)-P_{ij}(-h_i)=
h_iM_\pi(M^2_i-m^2_j)|\bq|
\,,
}
while the sum is given by Eq.\,\eqref{p=q+k}. One can see this results
are in accordance with Eq.\,(2.38) of Ref.\,\onlinecite{shrock2} where the
polarization $[P_{ij}(+)-P_{ij}(-)]/[P_{ij}(+)+P_{ij}(-)]$ is calculated.

The ratio between the squared amplitudes for left-handed and right-handed
helicities is
\eq{
\label{ratio}
R_{ij}=\frac{P_{ij}(h_i=-1)}{P_{ij}(h_i=1)}=
\frac{m^2_j}{M^2_i}\frac{M^4_\pi}{(M^2_\pi-M^2_i)^2}
\,.
}
Considering numerical values we obtain for $M_i=M_\mu$,
\eq{
R_{\mu j}=\frac{m^2_j}{(100\mathrm{keV})^2}\times 4.92\times 10^{-6}
\,,
}
while for $M_i=M_e$,
\eq{
R_{ej}=\frac{m^2_j}{(1\mathrm{eV})^2}\times 3.83\times 10^{-6}
\,.
}
Considering the actual direct bounds for the neutrino masses\,\cite{pdg}:
$m_{\nu_\mu}<190\mathrm{keV}$ and $m_{\nu_e}<2\mathrm{eV}$, we need a
precision of $10^{-6}$ in the helicity measurement to reach those bounds either
in the case of muons or electrons. Although the branching ratio to
produce muons from pion decay is much bigger than to produce electrons, the
required dominant versus wrong helicity probability ratios are similar.

Therefore, the coefficient $\delta$ in Eq.\,\eqref{RR+LL} has exactly the
modulus
\eq{
\label{delta2}
|\delta_{\mu j}|^2=R_{\mu j} =
\frac{m^2_j}{M^2_\mu}\frac{M^4_\pi}{(M^2_\pi-M^2_\mu)^2}
\,.
}
If we rewrite
\eq{
|\delta_{\mu j}|=\frac{m_j}
{2E_{\nu_j}}\frac{M_\pi}{M_\mu}
\,,
}
where $E_{\nu_j}=\frac{M^2_\pi-M^2_\mu}{2M_\pi}+O(m^2_j)$, we see that
$|\delta_{\mu j}|$ is modified by the factor $\frac{M_\pi}{M_\mu}$ when
compared to the naive estimate $m_\nu/2E_\nu$\,\cite{langacker}. We can also
conclude that for the channel $\pi\rightarrow e\bar{\nu}$ the real factor is
enhanced considerably ($\sim 274\times$).

In general, experiments can not achieve a perfect accuracy in helicity
measurements because they usually involve polarization
distributions\,\cite{nuehelicity,roesch,fetscher}.
Thus we have to consider the accuracy necessary to be able to measure the
wrong, non-dominant helicity amplitude. 
Parametrizing Eq.\,\eqref{A^2:n} using $\ip{\hn}\hat{\bq}=-\cos\theta$ yields
\eqarr{
P_{ij}(\theta)&=&
M^2_i(M^2_\pi-M^2_i)\mn{\frac{1}{2}}(1-\cos\theta)
\\&&
+\ \frac{m^2_j}{2}
\Big\{
M^2_\pi+2M^2_i 
\cr&& +\ 
\cos\theta\big[M^2_\pi+\frac{2M^2_i}{M^2_\pi-M^2_i}\big]
\Big\}
\,.
}
For $\cos\theta=1$ ($\cos\theta=-1$) we recover the negative (positive)
helicity for the charged lepton.
Expanding around $\theta=0$, we obtain
\eq{
P_{ij}(\delta\theta)
\approx 
M^2_i(M^2_\pi-M^2_i)\frac{\delta\theta^2}{4}
+m^2_j\frac{M^4_\pi}{M^2_\pi-M^2_i}
\,.
}
Therefore we need an angular resolution of
\eq{
\delta\theta=2 \sqrt{R}\sim 10^{-3}
}
to be able to probe the ratio $R=R_{ij}$ in Eq.\,\eqref{ratio}.

Considering the leptonic mixing the measurement of the wrong helicity for
muons probes
\eq{
|\mathcal{M}(\pi\rightarrow \mu\bar{\nu}:h_\mu=-1)|^2=
|C|^2\frac{M^4_\pi}{M^2_\pi-M^2_\mu} (m^2_{\nu_\mu})_{\rm eff}
\,,~~
}
where $C\equiv 2G_FF_\pi V_{ud}$ and
\eq{
\label{mnumu}
(m^2_{\nu_\mu})_{\rm eff}\equiv
\sum_j|U_{\mu j}|m^2_j\,,
}
is an effective mass for the muon neutrino, analogous to
$m^2_\beta$\,\cite{giunti} inferred from the tritium beta decay experiments for
the electron neutrino. The effective electron neutrino mass $m^2_\beta$ is
defined as the expression in Eq.\,\eqref{mnumu} using $|U_{ej}|^2$ instead of
$|U_{\mu j}|^2$.
In fact, $m^2_\beta$ can be extracted from $\pi^-\rightarrow e^-\bar{\nu}$ by
measuring the electron with negative helicity.

To obtain the decay rate, we must multiply the amplitude squared by the factor 
\eq{
\frac{1}{4\pi}\frac{1}{2M_\pi}
\Big[\frac{v_{\nu_j}v_{l_i}}{v_{\nu_j}+v_{l_i}}\Big]
\approx
\frac{1}{4\pi}\frac{1}{4M_\pi}
\Big(1-\frac{M^2_i}{M^2_\pi}\Big)
+ O(m^2_\nu)
\,,
}
arising from the phase space.
We then obtain
\eqarr{
\label{Gamma+}
\Gamma_\alpha\ms{(h_\alpha\!=\!1)}
\!\!&=&\!\!
\frac{G^2_F}{4\pi}F^2_\pi|V_{ud}|^2M_\pi
M^2_{\alpha}\Big(1-\!\frac{M^2_\alpha}{M^2_\pi}\Big)^2
\!\!+O(m^2_\nu)
,\quad~~
\\
\label{Gamma-}
\Gamma_\alpha\ms{(h_\alpha\!=\!-1)}
\!\!&=&\!\!
\frac{G^2_F}{4\pi}F^2_\pi|V_{ud}|^2M_\pi(m^2_{\nu_\alpha})_{\rm eff}
+O(m^4_\nu)
\,,
}
where $\alpha=e,\mu$ and $(m^2_{\nu_e})_{\rm eff}=m^2_\beta$.
The expression in Eq.\,\eqref{Gamma+} is the ordinary tree level decay rate for
the pion decaying into $l_\alpha\nu$\,\cite{DGH}.

\section{Discussions}
\label{sec:discussion}

The wrong helicity contribution come from the right-handed (helicity)
contribution in the left-handed (chirality) component of the neutrino which,
sometimes, can be explicitly decomposed as in the second term of
\eq{
\label{M+m}
\bar{u}_{l_i}(\bp)\sla{p}Lv_{\nu_j}(\bk)
=M_i\bar{u}_{l_i}(\bp)Lv_{\nu_j}(\bk)
-m_j\bar{u}_{l_i}(\bp)Rv_{\nu_j}(\bk)
,
}
where $p=q_i+k_j$ and Dirac equations are used.
Thus the wrong helicity contribution may receive new contributions from
operators containing the right-handed neutrino $\nu_R$.
Therefore, we can try to infer the presence of operators containing terms such
as 
\eq{
\label{nujR}
\bar{l}_iR\nu_j
}
coming from new interactions.
In the seesaw scenario, however, $\nu_{jR}\rightarrow N_{jR}$ would be heavy
Majorana fermions and their production is not possible. The production of
active neutrinos through active-sterile mixing is also suppressed since the MNS
matrix is unitary to a good extent\,\cite{antusch}.
The only scenario we can hopefully test here is the Dirac neutrino case with
neutrino masses and mixing coming from the Yukawa interactions in complete
analogy of the quark sector. The mystery of leptonic mass patterns and mixing
have to be explained by the same mechanism acting on the quark sector.

In the SM with three $\nu_{jR}$, without Majorana mass terms, there is no
physical interaction containing the term \eqref{nujR}. The simplest model
containing such term would be the Two-Higgs-doublet model (2HDM) which contains
the interaction with the physical charged scalar $h^{\pm}$:
\eq{
-\lag(l,\nu,h^{\pm})=
\Big[
(\Gamma^{e*}_1)_{ij}\bar{e}_{jR}\nu_{iL}
-(\Gamma^{\nu}_1)_{ij}\bar{e}_{iL}\nu_{jR}
\Big]h^-
+h.c.
}
The analogous interaction for quarks yields
\eq{
-\lag(Q,h^{\pm})=
\Big[
(\Gamma^{d}_1)_{ij}\bar{u}_{iL}d_{jR}
-(\Gamma^{u*}_1)_{ij}\bar{u}_{jR}\nu_{iL}
\Big]h^+
+h.c.
}
These interactions can be read off from the Yukawa interactions before
electroweak symmetry breaking (EWSB).
For quarks, for example, we have
\eqarr{
-\lag_Y(Q)&=&
(\Gamma^{\prime d}_a)_{ij}\bar{Q}_{iL}d_{jR}\Phi_a
\cr&&
+\ 
(\Gamma^{\prime u}_a)_{ij}\bar{Q}_{iL}u_{jR}\tilde{\Phi}_a
\,,~a=1,2
}
where $\tilde{\Phi}_a=i\sigma_2\Phi_a^*$ and 
\eqarr{
\Phi_1&=&
\begin{pmatrix}
h^+ \cr \frac{t_1-it_2}{\sqrt{2}}
\end{pmatrix}
\\
\Phi_2&=&
\begin{pmatrix}
G^+ \cr \frac{v-t_3+iG^0}{\sqrt{2}}
\end{pmatrix}
\,,
}
in the Higgs basis\,\cite{ccn:param}; $v=246\rm GeV$ is the electroweak 
vacuum expectation value.
After EWSB, we choose the basis for the fermionic fields such that 
$\Gamma_2^{d}=\frac{\sqrt{2}}{v}\diag(m_d,m_s,m_b)$,
$\Gamma_2^{u}=\frac{\sqrt{2}}{v}\diag(m_u,m_c,m_t)$, 
$\Gamma_2^{e}=\frac{\sqrt{2}}{v}\diag(M_e,M_\mu,M_\tau)$ and 
$\Gamma_2^{\nu}=\frac{\sqrt{2}}{v}\diag(m_1,m_2,m_3)$.

Using the relation\,\cite{DGH}
\eq{
\bra{0}\bar{u}\gamma_5d\ket{\pi^-(\bp)}=
-i\sqrt{2}F_\pi\frac{M^2_\pi}{m_u+m_d}
\,,
}
we obtain the amplitude for $\pi^-\rightarrow l_i^-\bar{\nu}_j$ for the
contribution coming from the exchange of $h^{\pm}$
\eq{
\label{troca:h}
-i\mathcal{M}(\pi^-\rightarrow l_i^-\bar{\nu}_j)=
-C'_{ij}
\bar{u}_{l_i}(\bq)Rv_{\nu_j}(\bk)
\,,
}
where
\eq{
C'_{ij}\equiv
\frac{(\Gamma^d_1+\Gamma^{u*}_1)_{11}}{m^2_{h^{\pm}}}
\frac{\sqrt{2}F_\pi M^2_\pi}{m_u+m_d}
(\Gamma^{\nu}_1)_{ij}
\,.
}

To quantify the contribution of Eq.\,\eqref{troca:h} we have to compare it to
the similar contribution coming from the second term of Eq.\,\eqref{M+m},
namely,
\eq{
\label{ratio:new/old}
\frac{|C'_{ij}|}{|m_jC|}=
\left|\frac{(\Gamma_1^d+\Gamma^u_1)_{11}(\Gamma^{\nu}_1)_{ij}}
{(m_u+m_d)V_{ud}}
\right|
\frac{M^2_\pi}{\sqrt{2}m^2_{h^{\pm}}G_Fm_j}
\,.
}
To estimate Eq.\,\eqref{ratio:new/old} we can assume the Yukawa couplings
$\Gamma^u_1$ e $\Gamma^d_1$ are of the same order as the couplings responsible
for giving the quark masses:
\eq{
(\Gamma^u_1)_{11}\sim (\Gamma^u_2)_{11}=\frac{\sqrt{2}}{v}m_u\,,~~
(\Gamma^d_1)_{11}\sim (\Gamma^d_2)_{11}=\frac{\sqrt{2}}{v}m_d\,.
}
Assuming the same for $\Gamma^\nu_1$, i.e.,
\eq{
(\Gamma^\nu_1)_{ij}\sim \frac{\sqrt{2}}{v}m_j\delta_{ij}\,,~~
}
we obtain
\eq{
\label{ratio:new/old:2}
\frac{|C'_{jj}|}{|m_jC|}\sim
10^{-3}\times \Big(\frac{100\text{GeV}}{m^2_{h^{\pm}}}\Big)^2
\,.
}
Since $m_{h^{\pm}}$ is unlikely to be smaller than 100GeV\,\cite{krawczyk},
the contribution from the physical charged Higgs exchange is generally
suppressed compared to the SM contribution unless the Yukawa couplings
$\Gamma^d_1,\Gamma^u_1$ or $\Gamma^\nu_1$ are much stronger than the ones
responsible for the respective masses.
Another natural possibility is 
\eq{
(\Gamma^\nu_1)_{ij}\sim \frac{\sqrt{2}}{v}m_iU'_{ij}\,,
}
and $m_i/m_j\sim 10^3$ or larger. Possibly, $U'_{ij}$ can be contrained by
flavor changing processes involving leptons\,\cite{casas-ibarra}.

To summarize, we can have access to the absolute neutrino mass scale beyond the
present direct search bounds if one can achieve a precision of $10^{-6}$ in the
measurement of the electron or muon helicity. In terms of polarization, an
angular resolution of $10^{-3}$ is necessary in the rest frame of the charged
particle. Although a large precision is required to perform such measurements,
it must be emphasized that the measurement should be performed only on the
charged lepton, without the need to detect the neutrinos directly.
For instance, another possible alternative method of accessing the absolute
neutrino mass scale would be the detection of flavor violating processes such 
as $\pi\rightarrow \mu\bar{\nu}_e$ which is also proportional to the square
of neutrino masses but depends on the detection of both charged lepton and
neutrino\,\cite{flavorLee}.
Moreover, contributions from the charged physical Higgs in 2HDMs are suppressed
compared to the SM contribution unless unnaturally large Yukawa couplings are
present. Nevertheless, the wrong helicity contribution for the channel
$\pi\rightarrow e\bar{\nu}_e$ is considerably enhanced if compared to naive
estimates. In this channel one can measure the same effective mass $m^2_\beta$
obtainable from the tritium beta decay experiments.
On the other hand, the analogous effective mass $(m^2_{\nu_\mu})_{\rm eff}$ may
be accessible in the dominant channel $\pi\rightarrow \mu\bar{\nu}_\mu$.
The alternative method of constraining new physics from the branching ratio
fraction Br($\pi\rightarrow e\nu$)/Br($\pi\rightarrow \mu\nu$) is investigate in
Ref.\,\cite{campbell}.



\acknowledgments
This work was 
supported by {\em Fundação de Amparo à Pesquisa do Estado de São Paulo} 
(Fapesp). The author is thankful to Prof. R. Shrock for pointing out
Refs.\,\cite{shrock,shrock2}.


\end{document}